   \documentstyle[twocolumn,prl,aps,psfig,epsf,amssymb]{revtex}
\begin{document}

%
% Defintions
%
\newcommand{\be}{\begin{equation}}
\newcommand{\ee}{\end{equation}}
\newcommand{\bea}{\begin{eqnarray}}
\newcommand{\eea}{\end{eqnarray}}
\newcommand{\beann}{\begin{eqnarray*}}
\newcommand{\eeann}{\end{eqnarray*}}
\newcommand{\bma}{\begin{array}{cc}}
\newcommand{\ema}{\end{array}}
\newcommand{\fr}{\frac}
\newcommand{\ra}{\rangle}
\newcommand{\la}{\langle}
\newcommand{\li}{\left}
\newcommand{\re}{\right}
\newcommand{\ri}{\right}
\newcommand{\uarr}{\uparrow}
\newcommand{\darr}{\downarrow}
\newcommand{\alp}{\alpha}
\newcommand{\df}{\stackrel{\rm def}{=}}
\newcommand{\nn}{\nonumber}
\newcommand{\dpl}{\displaystyle}
\def \gplus#1{{\cal G}^+ \left( #1 \right) }
\def \gminus#1{{\cal G}^- \left( #1 \right) }
\def \diff#1{{\cal D} \left( #1 \right) }
\def \lp {L_{\phi}}
\def \xo {x_1}
\def \xop {x_1^{\prime}}
\def \xt {x_2}
\def \yo {y_1}
\def \yop {y_1^{\prime}}
\def \yt {y_2}
%
% \draft command makes pacs numbers print
%
% \twocolumn[\hsize\textwidth\columnwidth\hsize\csname @twocolumnfalse\endcsname
\draft
\title{Quantum Transport in Nonuniform Magnetic Fields: \\
Aharonov-Bohm Ring as a Spin Switch
}

\author{Diego Frustaglia$^a$, Martina Hentschel$^a$, and Klaus Richter$^{a,b}$ }
\address{
$^a$ Max-Planck-Institut f\"{u}r Physik komplexer Systeme,
N\"othnitzer Str. 38, 01187~Dresden, Germany, \\
$^b$ Institut f\"ur Theoretische Physik, Universit\"at Regensburg,
D-93040 Regensburg, Germany}
\date{\today}
\maketitle
% \widetext
%
\begin{abstract}
{
We study the spin-dependent magneto conductance in mesoscopic
rings subject to an inhomogeneous in-plane magnetic field.
We show that the polarization direction of transmitted spin-polarized electrons
can be controlled via an additional magnetic flux such that spin flips
are induced at half a flux quantum.
This quantum interference effect is independent
of the strength of the nonuniform field applied.
We give an analytical explanation for one-dimensional rings
and numerical results for corresponding ballistic microstructures.
}
\end{abstract}
%

% \vspace*{6cm}

% to be published in  {\em ...}
%  \hspace*{4cm} MPI-PKS 01...

% \vspace*{-6.5cm}

\pacs{72.25.-b,05.30.Fk,73.21.-b,03.65.Bz}

% changed accroding to the editor's suggesutions

%03.65.Bz Foundations, theory of measurement, miscellaneous theories
%   (including Aharonov- Bohm effect, Bell inequalities, Berry's phase)
%03.65.Sq Semiclassical theory and applications
%05.45.+b theory and model of chaotic systems
%05.30.Fk Fermion system and electron gas
%72.25 Spin-polarized transport
%73.20.Dx Electron states in low dimensional structures
%
% ]

% \ifpreprintsty\tightenlines \else \begin{multicols}{2} \fi
\bibliographystyle{simpl1}
%\begin{multicols}{2}
%\narrowtext
%
%%%%%%%%%%%%%%%%%%%%%%%%%%%%%%%%%%%%%%%%%%%%%%%%%%%%%%%%%%%%%%%%%%%%
%%%%%%                  NEW SECTION
%%%%%%%%%%%%%%%%%%%%%%%%%%%%%%%%%%%%%%%%%%%%%%%%%%%%%%%%%%%%%%%%%%%%
%

Recent experimental progress \cite{spin-injection}
in creating spin-polarized charge carriers in semiconductors indicates 
the principle ability to perform spin electronics\cite{P98} based on
nonmagnetic semiconductors devices. This widens the field of
usual magneto-electronics in metals and opens up the intriguing
program of combining the rich physics of spin-polarized
particles with all the advantages of semiconductor fabrication
and technology, e.g.\ precise design of nanoelectronic
devices with controllable charge carrier densities and optoelectronical
applications. Besides, the spin relaxation times involved can be rather long;
coherence of spin-states can be maintained
up to scales of more than 100 $\mu$m\cite{KA99}. Hence coherent control and
quantum transport of spin states in semiconductor heterojunctions
or quantum dots is attracting increasing interest \cite{dassarma},
also in view of proposed future applications
including spin transistors\cite{DD90}, filters\cite{GB00}, and
scalable devices for quantum information processing\cite{LD98,IABDLSS99},
to name only a few.

In nonmagnetic semiconductors the coupling of the carrier spin to an applied
magnetic field can be used to control the spin degree of freedom. In
this respect, nonuniform magnetic fields whose direction varies on mesoscopic
length scales (textured fields)
are of particular interest. Besides the usual Zeeman spin splitting,
they give rise to a variety of additional effects absent in conventional
charge quantum transport.

In the limit of a strong magnetic field the electron spin can
adiabatically follow the spatially varying field direction,
and the spin wave function acquires a geometrical or Berry phase\cite{Ber84}.
In mesoscopic physics, Berry phases were first
theoretically studied for one-dimensional (1d) rings\cite{LGB90,Ste92}. 
They are also expected to give rise to clear signatures in
the magneto conductance of two-dimensional (2d) ballistic microstructures\cite{FR01}.
Nonuniform magnetic fields on mesoscopic scales have been realized in
semiconductors, for instance, by placing micromagnets\cite{YWGSKEN95,YTW98}
or ferromagnetic stripes\cite{NBH00} above or into the plane of a
2d electron gas in
high-mobility semiconductor heterostructures. However, although magnetic
inhomogeneities of up to 1 Tesla have been reported\cite{DGNLMH00},
it is difficult to experimentally reach the truly adiabatic regime.
The coupling of the carrier spins to more realistic moderate 
inhomogeneous fields generally lead to nonadiabatic, spin-flip 
processes counteracting geometrical phases.
Hence, despite various experimental efforts\cite{YTW98,MHKWB98} a clear-cut
demonstration of Berry phases in mesoscopic transport
remains an experimental challenge.

In this Letter we study {\em nonadiabatic}, spin-dependent coherent transport
through ballistic mesoscopic rings in the presence of textured fields. 
This enables us, on the one hand, to quantitatively 
investigate for {\em unpolarized} electrons the relevant conditions necessary
to observe geometrical phases or their nonadiabatic generalizations,
Aharonov-Anandan phases\cite{AA87}.
On the other hand
we show for {\em spin-polarized}
charge carriers how to use inhomogeneous fields to induce spin flips in a
controlled way. For Aharonov-Bohm (AB) ring geometries (with in-plane
nonuniform field) coupled symmetrically to two leads we demonstrate that
the spin direction of polarized particles transversing the rings can be tuned
and even reversed by applying an additional small perpendicular 
control field.
This quantum effect exists irrespective of adiabaticity.

%%%%%%%%%%%%%%%%%%%%%%%%%%%%%%%%%%%%%%%%%%%%%%%%%%%%%%%%%%%%%%%%%%%%%%%%%%%%%%%

\begin{figure}
\begin{center}
\centerline{\psfig{figure=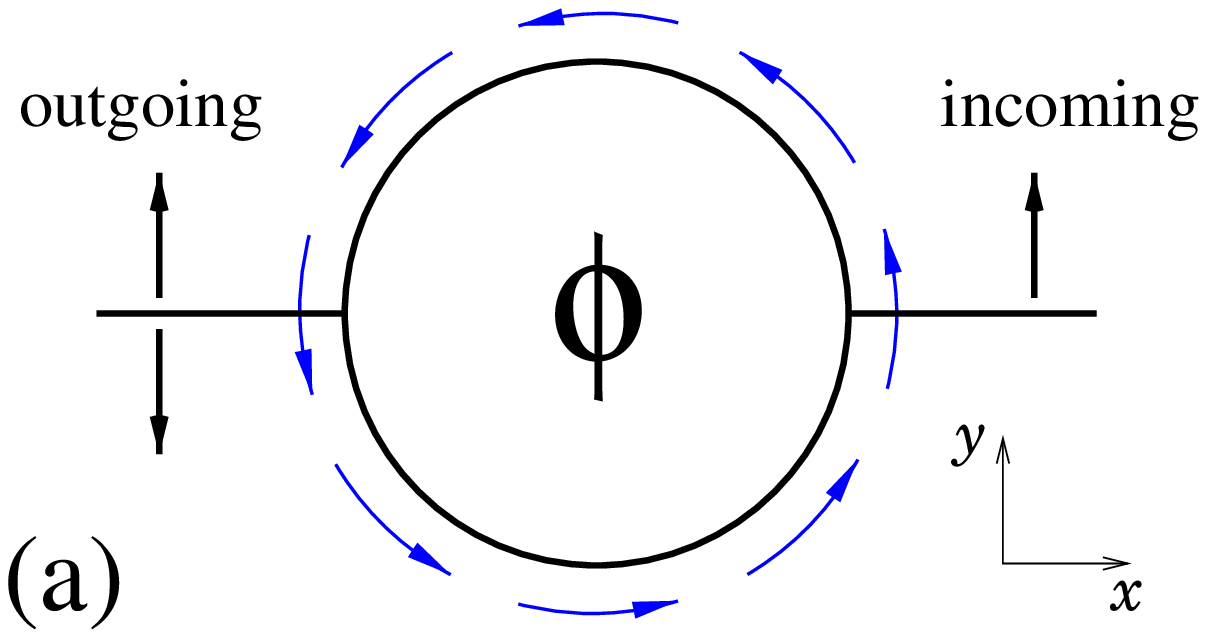,width=0.24\textwidth,angle=-90} 
\psfig{figure=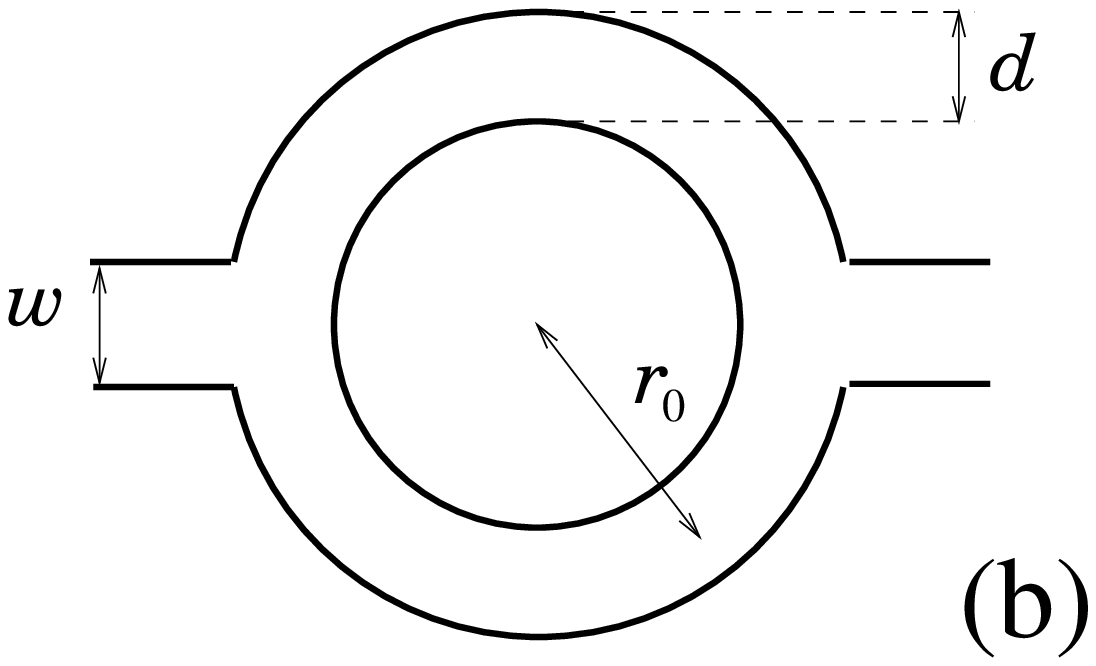,width=0.20\textwidth,angle=-90}}
\end{center}
\caption{
Geometries of ballistic microstructures used in the 
quantum calculations of the spin-dependent conductance for
a circular (in-plane) magnetic field texture plus a magnetic flux $\phi$. 
Spin directions are defined with respect to the $y$-axis. 
}
\label{fig1}
\end{figure}

We consider symmetric 1d and 2d ballistic mesoscopic rings with two
attached leads as shown in Fig.~\ref{fig1}.
For the inhomogeneous magnetic field we assume a
circular  configuration $\vec{B}_{\rm i}(\vec{r})= B_{\rm i}(r) \hat{\varphi}=  (a/r) \hat{\varphi}$
(in polar coordinates) centered around the inner disk of the microstructure.
Such a field can be viewed as being generated  by a perpendicular
electrical current through the disk\cite{current}.

The Hamiltonian for noninteracting electrons with effective mass
%$m^\ast$ and spin described by the Pauli spin matrix
$m^\ast$ and spin given by the Pauli matrix
vector $\vec{\sigma}$ reads, in the presence of a magnetic field
$\vec{B}=\vec{\nabla} \times \vec{A}$,
\be
\label{eq:qham}
H =  \fr{1}{2m^*}\li[\vec{p}+
    \fr{e}{c}\vec{A}(\vec{r})\re]^2+ V(\vec{r}) +
    \mu\ \vec{B} \cdot\vec{\sigma} \; .
\ee
The potential $V(\vec{r})$
defines the confinement of the ballistic conductor.
In our case the vector potential has two
contributions, $\vec{A}= \vec{A}_0+\vec{A}_{\rm i}$.
The term $\vec{A}_{\rm i}(\vec{r})$ generates the inhomogeneous field
$\vec{B}_{\rm i}(\vec{r})$ and $\vec{A}_0$ represents a (weak) perpendicular
uniform field $\vec{B}_0$ or an AB flux $\phi$
to be used as an additional
tunable parameter to study the magneto conductance.
In Eq.~(\ref{eq:qham}),  $\mu= g^* \mu_{\rm B}/2 =  g^* e \hbar/(4m_0\ c)$
with $\mu_{\rm B}$ the Bohr magneton,  $m_0$ the bare electron mass, $g^*$ the
effective gyromagnetic ratio, and $e >0$.

We compute the spin-dependent conductance $G(E,B_{\rm i},\phi)$ for two-terminal
quantum transport through the microstructures using the Landauer formula.
We focus on the case where the two leads of width $w$ 
% and the ring itself of width $d=w$ 
support only one open channel (Int$[k_{\rm F}w/\pi]=1$) 
\cite{one-channel}. The spin-dependent conductance then reads, for zero 
temperature, 
%Accounting for spin coupling to the inhomogeneous field
%the generalized expression for the conductance then reads,
%for zero temperature,
\be
\label{landauer}
G(E,B_{\rm i},\phi) = \frac{e^2}{h}
\li(|t^{\uarr\uarr}|^2+ |t^{\darr\darr}|^2 +
 |t^{\darr\uarr}|^2+|t^{\uarr\darr}|^2\re) \; .
\ee
We define the spin direction with respect to the  $y$ axis in 
Fig.~\ref{fig1}. The transmission coefficients
$T^{\darr\uarr}= |t^{\darr\uarr}|^2$  ($T^{\uarr\darr} = |t^{\uarr\darr}|^2$) describe
transitions between an incoming state from the right with spin up (down)
to an outgoing state to the left with spin down (up). They vanish for
$B_{\rm i} = 0$.
In the opposite, adiabatic limit of a strong magnetic field, the
magnetic moment associated with the electron spin 
travelling around the ring
stays (anti)aligned with the local inhomogeneous field. Hence, for the
field geometry in Fig.~\ref{fig1}(a) an incoming spin-up state is then converted
into a spin-down state upon transmission through the ring, and vice versa. In
the strong-field limit, 
$T^{\uarr\uarr}  = |t^{\uarr\uarr}|^2 = 0$ and $T^{\darr\darr} =
|t^{\darr\darr}|^2 = 0$.

For the experimentally relevant, intermediate case of moderate magnetic
fields one must solve coupled equations for the spin states
to account for spin flips.
We calculate the four spin-dependent transmission amplitudes by projecting the
corresponding Green function matrix of the system onto the transverse mode
spinors (of incoming and outgoing states) in the leads. We obtain the
Green functions for the Hamiltonian (\ref{eq:qham}) numerically after
generalizing the recursive Green function method for spinless
particles\cite{FG97} to the case with spin. This requires to replace the 
on-site and hopping energies in a tight-binding approach by 
$2 \times 2$ spin matrices.

We first study how adiabaticity is approached in mesoscopic rings
by considering the spin dependent transmission of unpolarized electrons
in the entire crossover regime between $B_{\rm i}\! =\! 0$ and the adiabatic limit. 
The appearance of geometrical phases requires an adiabatic separation
of time scales: For 1d rings of radius $r_0$ the Larmor frequency of spin precession,
$\omega_{\rm s}=2 \mu B/\hbar$, must be large compared to the 
frequency $\omega=v_{\rm F}/r_0$ of orbital motion
with Fermi velocity $v_{\rm F}$ around the ring\cite{Ste92}.
In the adiabatic limit a geometric phase  $\gamma^{\uarr(\darr)}$ is
acquired during a round trip. For the in-plane field configuration
considered,  $\gamma^{\uarr(\darr)} = \pi$ giving rise to a
geometric flux $-\phi_0/2$ with   $\phi_0 = hc/e$.
Together with an AB flux $\phi$ this adds up to an
effective flux $\phi - \phi_0/2$
and leads to a shift in the AB
magneto oscillations of $T^{\uarr\darr}$ and $T^{\darr\uarr}$
such that the overall transmission $T(\phi\! =\!0) = 0$ \cite{FR01},
since also $T^{\uarr\uarr}$ and $T^{\darr\darr}$ tend to zero, see above.

The  condition for adiabaticity can be written as
\be
\label{1Dadiab}
q  \equiv \fr{\omega}{\omega_{\rm s}} =
  \frac{k_{\rm F} r_0}{g^\ast (m^\ast/m_0) (\pi r_0^2B/\phi_0)} \ll 1 \; ,
\ee
with $k_{\rm F} = m^\ast v_{\rm F}/\hbar$.
For 2d rings of width $d$ and mean radius $r_0$ the angular ($\hat{\varphi}$)
component of $\vec{k}_{\rm F}$ is relevant for adiabaticity. For the
$m$-th propagating mode in a 2d ring, $q$ in  Eq.~(\ref{1Dadiab}) is then
replaced by the rescaled parameter
$q_{\varphi} \equiv q \sqrt{1 - [m/(k_{\rm F}d/\pi)]^2}$ (provided that
$d/r_0 \ll 1$).

\begin{figure}
\begin{center}
\centerline{ \psfig{figure=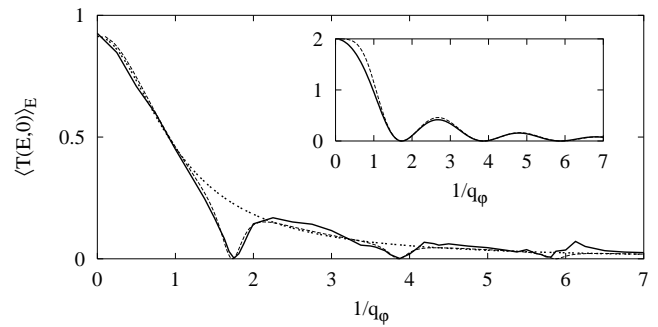,width=0.50\textwidth,angle=-90}}
\end{center}
\vspace*{-4mm}

\caption{Energy-averaged quantum transmission as
a function of the adiabaticity parameter $q_{\varphi}$ (see text)
for unpolarized incoming electrons through rings with circular
in-plane magnetic field  (as in Fig.~\ref{fig1}) and zero flux. 
The solid line represents numerical results for the geometry
in  Fig.~\ref{fig1}(b). The dashed curve  shows results from
a corresponding transfer-matrix approach for a 1d ring (Fig.~\ref{fig1}(a))
with coupling $\epsilon=0.316$ (see text).
The dotted line shows an overall Lorentzian 
dependence $0.916/(1+q_\varphi^{-2})$. Inset: 1d approximate 
(Eq.~(\ref{1d-meanT}), dashed) and  full (solid) results for strongly 
coupled leads ($\epsilon=0.5$). 
In all curves $\langle T(E,\phi=0) \rangle_E \rightarrow 0$ 
for $q_\varphi \rightarrow 0$ as a result of geometrical phases.
}
\label{fig2}
\end{figure}

To show how adiabaticity is approached we consider transport 
through a ring with one open channel, $m\!=\!1$.
The solid line in Fig.~\ref{fig2} depicts the numerically obtained
average transmission $\langle T(E,\phi\!=\!0)\rangle_E$  
as a function of $1/q_{\varphi}$
for the quasi-1d ring of Fig.~\ref{fig1}(b) ($d/r_0\!=\!0.25$) at $\phi
\!=\!0$.
The average is taken over an energy interval (between the
first and second open channel) at fixed $q_{\varphi}$ to smooth out
energy-dependent oscillations. With increasing $1/q_{\varphi}$ the
transmission $\langle T(E,\phi=0)\rangle_E$ 
tends to zero which is a clear signature of the 
geometrical phase as discussed above. The overall decay
is Lorentzian, $\sim(1+q_{\varphi}^{-2})^{-1}$ (dotted
line in Fig.~\ref{fig2}). This curve and the dashed
line, which well agrees with the numerical result,
is obtained in an independent transfer matrix approach for
a 1d ring  (Fig.~\ref{fig1}(a), $q_{\varphi} \equiv q$)
to be discussed below. 

In our numerical calculations $\langle  k_{\rm F} r_0
\rangle \simeq 15$. In a typical experimental setup, $ k_{\rm F} r_0 = 2\pi
r_0/\lambda_F \simeq 60$ for $r_0 \simeq 500$ nm. Then we have $1/q \simeq$
0.07$ B$[T] and 0.86$ B$[T] for GaAs and InAs.
However, despite the relatively large fields
necessary for satisfying Eq.~(\ref{1Dadiab}) for $q$, 
the scaling factor entering into $q_{\varphi}$
allows one to reach adiabaticity for considerably
lower field strengths. This can be achieved either
by reducing the width $d$ of quasi-1d rings
or by reducing $k_{\rm F}$ by variation of the electron density 
\cite{pedersen}.

%%%%%%%%%%%%%%%%%%%%%%%%%%%%%%%%%%%%%%%%%%%%%%%%%%%%%%%%%%%%%%%%%%%%%%%%%%

\begin{figure}
\begin{center}
\centerline{ \psfig{figure=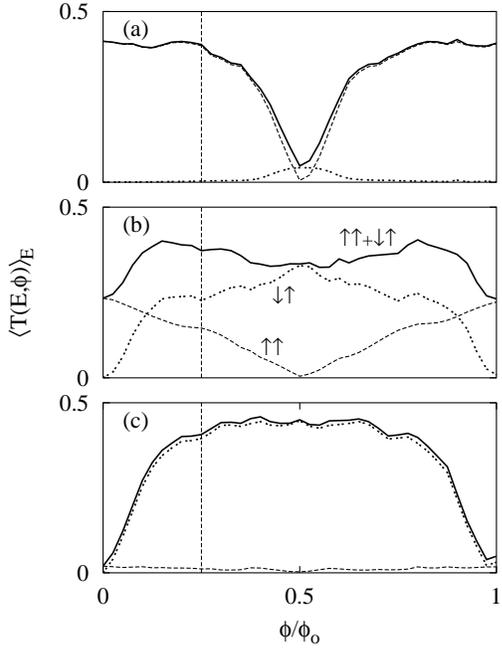,width=0.40\textwidth,angle=0}}
\end{center}
\vspace*{-5mm}

\caption{Averaged transmission for up-polarized incoming electrons 
(see Fig.~1) through
a quasi-1d ring as  function of a flux $\phi=\pi r_0^2 B_0$
in the presence of a circular in-plane field $B_{\rm i} \gg B_0$ 
of increasing strength: (a) weak, (b) moderate, (c) strong.
The overall transmission (solid line) is split into its components 
$\langle T^{\uparrow \uparrow} \rangle$ (dashed) and 
$\langle T^{\downarrow \uparrow}\rangle$ (dotted). 
Note the change in the polarization upon tuning the flux
and the spin-switch mechanism at $\phi = \phi_0/2$. 
}
\label{fig3}
\end{figure}

In the following we study how the spin-dependent transmission changes
as a function of an additional flux $\phi = \pi r_0^2 B_0$ with $B_0\! \ll\!
B_{\rm i}$. 
Our main results are summarized in
Fig.~\ref{fig3}(a)-(c), which shows the average  $\langle
T(E,\phi)\rangle_E$ for three different scaled strengths 
$q_\varphi \approx$ 20, 1.4,
0.25 of the inhomogeneous field. 
We consider up-polarized, incoming spins; equivalent results are obtained for
spin-down states. In the weak-field limit, Fig.~\ref{fig3}(a), 
the coefficient $\langle T^{\darr\uarr}\rangle$ 
(dotted line) is close to zero, and the
total transmission (solid line) shows usual AB oscillations predominantly
given by $\langle T^{\uarr\uarr}\rangle$ (dashed line). 
The behaviour is reversed in the adiabatic limit,
panel (c), where  $\langle T^{\darr\uarr}\rangle$ 
exhibits AB oscillations, 
shifted by $\phi_0/2$ due to the geometrical phase as discussed above.

Panel (b) shows the general case of an intermediate
field. With increasing flux the polarization of transmitted electrons
changes continuously. Most interestingly, $\langle T^{\darr\uarr} \rangle=0$
at $\phi = 0$, while $\langle T^{\uarr\uarr} \rangle = 0$ for
$\phi = \phi_0/2$. For zero flux an ensemble of
spin-polarized charge carriers is transmitted always keeping
the spin direction, while for $\phi =\phi_0/2$ the transmitted
electrons just reverse their spin direction. In other words, 
by tuning the flux from $0$ to $\phi_0/2$, one can 
reverse the polarization of transmitted particles in a controlled way.
Hence, the AB ring plus the rotationally symmetric magnetic
field acts as a tunable spin switch, {\em independent} of the field strength
$B_{\rm i} > 0$,
which determines only the size of the spin-reversed current.
Alternatively, for a fixed flux $0 < \phi < \phi_0/2$
(vertical dotted line in Fig.~\ref{fig3})
the spin polarization is reversed upon going from the 
nonadiabatic to the adiabatic regime,  while 
the total transmission remains nearly constant.

This mechanism for changing the spin direction does 
neither rely on the spin coupling to the control field $B_0$, nor
on the  Zeeman splitting 
often exploited in spin filters. It is a pure quantum
interference effect which exists
also for the transmission at a given Fermi energy. 

In the following we give an analytical explanation for the
numerically observed effects (Figs.~\ref{fig2},\ref{fig3}). 
To this end we consider the 
model of a 1d AB ring coupled to 1d leads,  Fig.~\ref{fig1}(a),
and extend the transfer matrix approach for spinless particles\cite{imry} 
to the case with spin.  We follow the method outlined
in \cite{yi} but consider fluxes instead of probabilities  to
work with unitary transfer matrices.

The eigenstates of the Hamiltonian
(\ref{eq:qham}), which are  analytically obtained
for a ballistic 1d ring\cite{Ste92}, are necessary for 
implementing the transfer matrix algorithm.
They read $\Psi_{n,s} = {\rm exp}(i n \varphi)
\otimes \psi_n^s(\varphi)$ where the first factor describes the motion
along the ring and the second refers to the spin state $s=\uarr,\darr$
(with respect to the vertical $(z)$ axis).
The Zeeman term causes a slight difference in
the kinetic energy of spin-$\uparrow$ and spin-$\downarrow$
electrons travelling clockwise or  
counter-clockwise around the ring so that we must
distinguish four possible $n$: $n_j^{\uparrow},
n_j^{\downarrow}$  $(j\!=\!1,2)$.
They are given by $n' \equiv n + \phi/\phi_0$ where
the $n'$ are the solutions of the equation\cite{inprep}
$ \tilde{E_{\rm F}} = 
n'^4 + 2 n'^3 + ( 1 - 2 \tilde{E_{\rm F}}) n'^2 -
    2 ( \tilde{E_{\rm F}} +  \tilde{\mu} B \cos \alp) n' +
      \tilde{E_{\rm F}}^2 -  \tilde{\mu} B \cos \alp -
        ( \tilde{\mu} B)^2
$.
Here, $\tilde{E_{\rm F}} = (2m^* r_0^2/\hbar^2)~E_{\rm F}$ is
the scaled Fermi energy,
$ \tilde{\mu}=(2m^* r_0^2/\hbar^2)~\mu$, and $\alpha$  the tilt angle
of the textured magnetic field with respect to the $z$-axis.
In the general, nonadiabatic case four angles 
$\gamma_j^{\uparrow}, \gamma_j^{\downarrow} \leq \alpha$ 
take the r\^ole of $\alpha$ 
and characterize the spin eigenstates which read 
\[
\begin{array}{ccc}
    \psi^\uarr_{n_j} = %e^{i n_j^{\uarr} \varphi}
                    \li( \begin{array}{c}
                         \cos \fr{\gamma_j^{\uarr}}{2}
                         \\
                          \pm i e^{i \varphi} \sin
                         \fr{\gamma_j^{\uarr}}{2}
                         \end{array}
                         \re)
    & , &
    \psi^\darr_{n_j} = %e^{i n_j^{\darr} \varphi}
                    \li( \begin{array}{c}
                         \sin \fr{\gamma_j^{\darr}}{2}
                           \\
                         \mp i e^{i \varphi'} \cos
                         \fr{\gamma_j^{\darr}}{2}
                         \end{array}
                         \re) , 
  \end{array} 
\]
with $   \cot\gamma_j  =
\pm [\cot\alp + (2n'_j+1)/ (2\tilde{\mu} B \sin\alp)] $.
In the adiabatic limit $\gamma \rightarrow \alpha$.
The transfer matrices in the generalized basis 
enter into the transmission formulae 
which generally require for numerical evaluation.

For an in-plane  field
field $(\alpha = \pi/2)$ without additional flux ($ n \!=\! n'$)
the equations above simplify considerably and we find  
$ n_1^{\uarr(\darr)} \!=\! -(n_2^{\darr(\uarr)} \! + \! 1)$.
Though being involved, all expressions leading to the
transmission can be handled analytically.
The transmission  depends strongly
on the coupling at the junctions 
between the ring and the leads. It is given by a parameter 
$\epsilon$ \cite{imry} where 
$\epsilon=0$ $(0.5)$ describes zero (strongest) 
 coupling.  Adjusting  $\epsilon$
 the analytical model allows us to estimate the 
effective coupling to the leads in ballistic rings used in the 
numerical calculations above.
The dashed line in Fig.~\ref{fig2}  
($\epsilon=0.316$) fits well with the numerical result (solid).
For $\epsilon = 0.5 $ 
and $\phi=0$, an approximate analytical 
expression for  
$\langle T(E,0)\rangle_E$ can be given in compact form, if
we replace the energy averages over rapidly oscillating angular
functions involved by their mean. We find,
leaving the details to\cite{inprep}, 
\be
\langle T(E,\phi=0)\rangle_E
\simeq 16  \fr{\cos^2 \bar{\gamma_1} \sin^2 (\Delta n \pi/2) }
                        {4 + \cos^4 \bar{\gamma_1} \li[ 1 - \cos (\Delta n \pi) \re]^2}
                        \: .
\label{1d-meanT}
\ee
$\Delta n \equiv  n_1^{\uarr} - n_1^{\darr}$ and
$\bar{\gamma_1} \equiv (1/2) (\gamma_1^{\uarr} + \gamma_1^{\darr})$
can be expressed through $q_\varphi$
as $\Delta n = (1+q_\varphi^{-2})^{1/2}$ and $\cos \bar{\gamma_1} = 
(1+q_\varphi^{-2})^{-1/2}$.
The inset of Fig.~\ref{fig2} shows the result (\ref{1d-meanT})
(dotted line) compared to the exact 1d result (solid) 
for $\epsilon\!=\!0.5$. 

All the general features of $\langle T(E,0)\rangle_E$ in Fig.~\ref{fig2}
are well described by Eq.~(\ref{1d-meanT}). Owing to 
destructive interference,
the transmission vanishes at points where $\Delta n$ is an even integer 
corresponding to $1/q_\varphi= \sqrt{3}, \sqrt{15}, \ldots $ . 
Eq.~(\ref{1d-meanT}) gives a complicated overall decay factor for
$\langle T(E,0)\rangle_E$ which reduces to the Lorentzian $\cos^2
\bar{\gamma_1} = 1/(1+q_\varphi^{-2})$ 
in the limit $\epsilon \rightarrow 0$.
Already for $\epsilon < 0.4 $ this is
a good approximation for the overall crossover from the diabatic to the
adiabatic regime (dotted line in Fig.~\ref{fig2}).

Within the 1d model we further reproduce the flux-dependence for
spin-dependent transport, Fig.~3, and find 
an analytical proof\cite{inprep} for the spin switch effect 
discussed above.  We can show that
the transmission coefficient $T^{\uarr\uarr}$ vanishes
completely at $\phi = \phi_0/2$, if 
the magnetic field to which the spins couple 
has no component perpendicular to
the plane of the ring. 
Moreover, numerical evidence (not presented here) suggests 
that this condition might not be necessary. Furthermore, we find
numerically that the spin-switch effect does not only occur
in quasi-1d rings but persists, much more generally, in
doubly connected mesoscopic structures with more than one open mode
and arbitrary shape, as long as reflection
symmetry with respect to the horizontal axis is preserved\cite{inprep}. 
However, the effect requires single-channel leads\cite{one-channel}.
We further note that rings with Rashba (spin-orbit) interaction\cite{Nitta}
(yielding an effective in-plane magnetic field in the presence of 
a vertical electric field) might lead to a similar spin-switch effect. 

%%%%%%%%%%%%%%%%%%%%%%%%%%%%%%%%%%%%%%%%%%%%%%%%%%%%%%%%%%%%%%%%%%%%

To summarize, 
we have studied {\it nonadiabatic} spin transport through ring geometries 
subject to inhomogeneous magnetic fields.
We obtain, both numerically and analytically, the explicit dependence of the
transmission on the scaled field strength $q_\varphi$, which acts as an
adiabaticity parameter, elucidating the r\^ole of geometrical phases 
in quantum transport and possible experimental realizations.
For in-plane field geometries and  symmetric ballistic microstructures
 we demonstrate how an additional
small flux $\phi$ can be used to control spin flips and  to tune
the polarization of transmitted electrons.
This quantum mechanism does not require adiabaticity. 
In combination with a spin detector such a device may be used to control spin 
polarized current, similar to the spin field-effect transistor 
proposed in \cite{DD90}.
For ferromagnetic (generally diffusive) conductors 
disorder will break the spatial symmetry. However, the question, whether a related 
effect may prevail when considering disorder-averaged quantities, remains as 
a further interesting problem.

% \acknowledgments
We thank H.~Schomerus for very helpful comments and
J.~Fabian and D.\ Weiss for useful discussions.

%%%%%%%%%%%%%%%%%%%%%%%%%%%%%%%%%%%%%%%%%%%%%%%%%%%%%%%%%%%%%%%%%%%%

%%%%%%%%%%%%%%%%%%%%%%%%%%%%%%%%%%%%%%%%%%%%%%%%%%%%%%

%%%%%%%%%%%%%%%%%%%%%%%%%%%%%%%%%%%%%%%%%%%%%%%%%%%%%%

%%%%%%%%%%%%%%%%%%%%%%%%%%%%%%%%%%%%%%%%%%%%%%%%%%%%%

%%%%%%%%%%%%%%%%%%%%%%%%%%%%%%%%%%%%%%%%%%%%%%%%%%%%%%%


\begin{references}
\bibitem{one-channel}
This is achieved in semiconductor heterostructures by tuning the coupling
of the ring to the leads through additional gate voltages such
that only one channel is open.
\end{references}
\end{document}